\documentclass[aps,superscriptaddress,twocolumn,floatfix]{revtex4-1}  

\usepackage{hyperref}
\usepackage{color}
\usepackage{amsmath,amsfonts,amssymb,amsthm,graphics,graphicx,epsfig,times,bbm}
\usepackage[dvipsnames]{xcolor}
\usepackage[mathscr]{euscript}
\usepackage[sort&compress]{natbib}
\usepackage{scalefnt}

\newcommand\ket[1]		{\lvert#1\rangle\mspace{-1.5mu}}		
\newcommand\bra[1]		{\mspace{-1.5mu}\langle#1\rvert}		

\newcommand\ens[1]		{\left\{#1\right\}}										

\newcommand\ox{\otimes}
\newcommand\herm{^\dagger}

\newcommand\sox[1]{^{\ox #1}}

\newtheoremstyle{PRLthm}
  {5pt}
  {5pt}
  {\itshape}
  {}
  {\bfseries}
  {:}
  { }
  {\thmname{#1}\thmnumber{~#2}\thmnote{ \textmd{(#3)}}}
  
\theoremstyle{PRLthm}
\newtheorem{obs}{Observation}

\newcommand\cH[1][2]{\mathcal H_{#1}}

\newcommand\union{\cup}

\DeclareMathOperator\Symm{Symm}

\DeclareMathOperator\supp{supp}
\newcommand\idop{\openone}

\def\unqm2sat{\textsc{unique quantum 2-sat}}
\def\qm#1sat{\textsc{quantum ${#1}$-sat}}

\usepackage{color}

\def\id{\mathbbm{1}}

\begin{document}

\title{Solving frustration-free spin systems}

\author{N.\ de Beaudrap}
\author{M.\ Ohliger}
\affiliation{Institute of Physics and Astronomy, University of Potsdam, 14476 Potsdam, Germany}

\author{T.\ J.\ Osborne}

\affiliation{Institute for Advanced Study Berlin, 14193 Berlin, Germany}

\author{J.\ Eisert} 

\affiliation{Institute of Physics and Astronomy, University of Potsdam, 14476 Potsdam, Germany}
\affiliation{Institute for Advanced Study Berlin, 14193 Berlin, Germany}

\date{\today}

\vspace*{-.6cm}

\begin{abstract}
We identify a large class of quantum many-body systems that can be solved exactly: natural frustration-free 
spin-$1/2$ nearest-neighbor Hamiltonians on arbitrary lattices. We show that the entire ground state 
manifold of such models can be found exactly by a tensor network of isometries acting on a space 
locally isomorphic to the symmetric subspace. 
Thus, for this wide class of models real-space renormalization can be made exact. 
Our findings also imply that every such frustration-free spin model satisfies an area law for the
entanglement entropy of the ground state, establishing a novel large class
of models for which an area law is known. Finally, we show that our approach
gives rise to an ansatz class useful for the simulation of almost
frustration-free models in a simple fashion, outperforming mean field
theory.
\end{abstract}
\maketitle

Understanding the physics of quantum many-body systems is a central goal of modern physics, as they can exhibit exotic phenomena with no parallel in classical physics, including topological effects and quantum phase transitions at zero temperature.
However, the very source of their rich physics also leads to a major roadblock in their study: the Hilbert space dimension of these systems scale exponentially with the number of particles.
This means that brute-force numerical techniques fail even for systems of only a handful of particles.  

A key insight in the study of local quantum many-body systems is that naturally occurring states only occupy a small subspace of the Hilbert space which in principle is available to them.
Specifically, it has been realized that ground, thermal, and dynamically evolving states are only weakly entangled: the entanglement entropy satisfies what is referred to as an \emph{``area law''}~\cite{Area1,Area2,Area3}.
This insight is the basis of the density-matrix renormalization group approach, and higher-dimensional analogues \cite{Scholl}.
So successful are these methods in practice that one is tempted to boldly conjecture that all \emph{physically} relevant systems will soon be tractable to one or another of the numerical tools we have to hand.
Recent important results give cause for caution, showing that general numerical methods cannot well-approximate the physics of an \emph{arbitrary} local quantum system, even in 1D: these include local glassy models where approximating the ground state energy is \emph{QMA-complete}~\cite{Power}, suggesting that such systems would be intractable even for quantum computers.
However, these results do not give much reason for practical concern thus far, as at least in 1D they rely on rather baroque constructions (involving very large local dimensions).

In this work, we approach the issue of the complexity or ``hardness'' of finding ground states from the other direction:
We establish a large class of models for which the task of finding the ground state is \emph{easy}, in that the ground state manifold can be described exactly and efficiently. This is the class of \emph{all natural frustration free spin-$1/2$ models with 
nearest-neighbor interaction on general lattices}.
(By the qualifier ``natural'', we mean that all two-spin interaction terms have excited states which are entangled, which might be taken as implicit in the semantics of an ``interaction term''.)
Adopting and extending ideas of Ref.~\cite{Bravyi06} on \qm2sat\, and going beyond 1D models as in Ref.~\cite{Farhi},
we find that the complete 
ground state manifold of such Hamiltonians can be constructed by considering the ground spaces of each interaction term in turn, and obtaining a reduction to the symmetric subspace of a smaller system.
In doing so, we find that the resulting ground state manifold can be efficiently grasped in terms of tree-tensor networks.
We discuss how this allows expectation values of local observables to be computed efficiently. 
What is more, the ground states satisfy an {\it area law}.
Physically, we can view this work as describing a large class of models for which an instance of {\it real-space renormalization} 
provides an \emph{exact} solution to the true genuine quantum many-body model. Finally, we see how this construction --- a tree tensor network with a symmetric subspace as an input --- can serve as an ansatz class to simulate \emph{almost frustration-free models}
which are in a sense ``close'' to frustration-free models.
In this way, one can outperform mean-field approaches in a very simple fashion.

In our analysis, we allow for nearest-neighbor Hamiltonians on arbitrary lattices.
This could be a cubic lattice of some dimension, or more generally any graph, the vertex set of which we denote by $V$.
On this lattice, the spin Hamiltonian $H$ is represented as
\begin{gather}
	\label{eqn:inputHamiltonian}
		 		H = \sum_{\{a,b\}} h_{a,b}
\end{gather}
for terms $h_{a,b}$ acting on pairs of spins $\{a,b\} \subset V$.
By rescaling, we may without loss of generality require that the ground energy of each term $h_{a,b}$ is zero.

The ground state manifold $M$ of such a Hamiltonian may be degenerate: we identify the ground state $\rho$ with the maximal mixture over $M$.
(This is a pure state only if $H$ is non-degenerate.)
We describe properties of the ground state $\rho$, and more generally the manifold $M$, given $H$ as in Eq.~\eqref{eqn:inputHamiltonian}.
The Hamiltonian $H$ is \emph{frustration-free} (or \emph{unfrustrated}) if the ground state vectors $\ket{\Phi} \in M$ are ground states of individual coupling terms, that is if $h_{a,b}\ket{\Phi} =0$ holds for all  $h_{a,b}$ and all $|\Phi\rangle \in M$; we say otherwise that $H$ is \emph{frustrated}.

We will call a spin Hamiltonian $H$ \emph{natural} if it contains no isolated subsystems, and each interaction term $h_{a,b}$ (considered as an operator on $\mathbb C^2 \ox \mathbb C^2$) has at least one entangled excited state (i.e., an entangled state orthogonal to the ground state manifold of $h_{a,b}$).
In what follows we will consider only such ``natural'' Hamiltonians.

\textit{Frustration-free spin Hamiltonians.}
Our main results concern the class of (natural) frustration-free spin Hamiltonians $H$.
We show that the ground state manifold $M$ of a frustration-free spin Hamiltonian on $N$ spins has dimension at most $N+1$.
What is more, it is the image of a space of low Schmidt measure~\cite{SM} 
under a tree-tensor network.
We proceed by extending the work of Ref.~\cite{Bravyi06} for \qm2sat.

First, we describe the needed components of Ref.~\cite{Bravyi06}, in the language of frustration-free models.
\label{discn:reviewBravyi}
Consider a Hamiltonian $H_U$ containing terms $h_{u,v }$ of rank $2$ or $3$. 
If $H_U$ is frustration-free, the reduced state $\rho_{u,v}$ of any state vector $\ket{\Phi} \in \ker(H)$ is in the kernel of $h_{u,v}$\,; we may then consider a subspace $S_{u,v} \subset \cH[u] \otimes \cH[v]$ of dimension $2$ which contains $\supp(\rho_{u,v})$.
By defining an isometry
\begin{align}
		R_{u:uv} : \cH\sox{\ens{u}} \longrightarrow  S_{u,v} \subset \cH\sox{\ens{u,v}},
\end{align}
we can reduce to a Hamiltonian on fewer spins: we let
\begin{align}
	\label{eqn:isometricContraction}
			H'_U
		\;=\;
			R_{u:uv}\herm H_U R_{u:uv}
		\;=\;
			\sum_{\{a,b\}}	R_{u:uv}\herm h_{a,b} R_{u:uv}	\,.
\end{align}
Such a spin Hamiltonian $H'_U$ is a sum of two-spin interactions (and possibly single-spin terms) of the form $h'_{a,b} = R_{u:uv}\herm h_{a,b} R_{u:uv}$.
(If $h_{u,v}$ has rank 3, then $h'_{u,v}$ is a non-zero single spin operator acting on $u$ alone.)
If $H$ contains non-zero terms $h_{a,u}$ and $h_{a,v}$, we obtain two non-zero contributions $h'_{a,u} = R\herm_{u:uv} h_{a,u} R_{u:uv}$ and $h'_{a,v} = R\herm_{u:uv}  h_{a,v}  R_{u:uv}$ in the Hamiltonian $H'$, each of which act on $\{u,a\}$.
The sum gives a combined term $
		\bar h'_{a,u} = h'_{a,u} + h'_{a,v} 
$ 
in $H'_U$, possibly of higher rank than either $h'_{a,u}$ or $h'_{a,v}$~\cite{Note0}.
If the new Hamiltonian $H'_U$ contains terms of rank $2$ or $3$, we may perform another such reduction, and so on.
This reduction procedure has the following features:

\paragraph{Preservation of the kernel dimension.}
\label{remark:isometriesTreeTensorNetwork}
By construction, we have $\ker(H_U) = R_{u:uv} \ker(H'_U)$.
Thus, the kernels of $H_U$ and $H'_U$ have the same dimension.
In particular, if $H'_U$ has any terms of full rank acting either on one or two spins, then $\dim \ker(H_U) = 0$, in which case $H_U$ is frustrated.
If no full-rank terms are produced, each reduction leads to an operator acting on fewer spins, until we obtain a Hamiltonian having only terms of rank $1$.

\paragraph{Arbitrariness of reduction order.}
Because the dimension of the kernel is preserved by these reductions, we may continue to perform such reductions until we obtain a Hamiltonian which either (\textit{i})~contains only terms of rank 1, or (\textit{ii})~contains a full-rank term.
The latter cannot occur in the reduction of a frustration-free Hamiltonian; and we discuss below the analysis for Hamiltonians having only rank-1 terms.
Thus, we may choose any convenient reduction sequence.

The above features allow us to reduce to the special case of a Hamiltonian $H'_U$ (acting on a system $V'$) which has only interaction terms of rank 1.
Each two-spin Hamiltonian term $h'_{a,b} = \ket{\beta_{a,b}}\bra{\beta_{a,b}}$ may be regarded as imposing constraints on the corresponding two-spin marginals
$\rho_{a,b}$ of states $\ket{\Phi} \in \ker(H)$: we aim 
to obtain additional constraints on pairs of spins $u,v \in V'$ by combining the known constraints.
To this end, Ref.~\cite{Bravyi06} shows that a state $\ket{\Phi}$ which is 
in the kernel of two functionals $\bra{\beta_{a,b}},\bra{\beta_{b,c}}$ is also in the kernel of
\begin{gather}
	\label{eqn:constraintInduction}
		\bra{\beta'_{a,c}}
	 = 
		\bigl( \bra{\beta_{a,b}} \ox \bra{\beta_{b,c}} \bigr) \bigl(  \id \ox \ket{\Psi^-} \ox  \id \bigr)
\end{gather}
acting on the spins $\{a,c\}$, where $\ket{\Psi^-}$ is the two-spin antisymmetric state vector.
For each such ``induced'' constraint $\bra{\smash{\beta'_{u,v}}}$ on spins $\{u,v\}$, we may add
the term $
	\tilde h_{u,v} = \ket{\beta'_{u,v}}\bra{\beta'_{u,v}}
$ 
to $H'_U$, resulting in a Hamiltonian $\tilde H_U$ which has the same kernel as $H'_U$.
(If $H'_U$ contains a term $h'_{u,v} \not\propto \tilde h_{u,v}$, it can be subsumed 
into a term $\bar h_{u,v}$ with rank at least $2$, in which case we apply a reduction $R_{u:uv}$ as above.)
One may induce further constraints from the terms of $\tilde H$, until we arrive at a \emph{``complete homogeneous''} Hamiltonian $H_c$, having only terms of rank 1, for which the constraints $\bra{\beta_{u,v}}$ are closed (up to scalars) under the constraint-induction procedure of Eq.~\eqref{eqn:constraintInduction}.

Ref.~\cite{Bravyi06} shows that such a Hamiltonian $H_c$, acting on at least one spin  and lacking 
single-spin operators~\cite{Note}, has a ground space containing product states.
Thus, the above remarks essentially recap the following result:

\begin{obs}[Ref.~\cite{Bravyi06}]
	There is an efficient algorithm to determine whether a spin Hamiltonian is frustration-free.
\end{obs}

We now extend the above results, to obtain a strong characterization of ground state manifolds for natural frustration-free systems.
We note the following three additional features of the isometric contraction scheme above:

\paragraph{Tree-tensor construction.}
The complete network of isometric reductions $T$ represents a \emph{tree-tensor network}, a special case of the \emph{MERA ansatz}~\cite{MERA} which is related to
real-space renormalization.
Each isometry $R_{u:uv}$ has one free input tensor index and two free output indices, and the sequential nature of the reduction ensures that the network is a directed and acyclic.
Thus, any spin $v$ introduced by an isometry $R_{u:uv}$ is a ``daughter spin'' of a unique parent $u$, leading to a tree-like structure on the tensor network $T$. 
Note however that $T$ has free input indices, corresponding to the roots of each tree: by construction, the ground space of $H_U$ is the image $T \ket{\Psi}$ of states $\ket{\Psi} \in \ker(H_c)$.

\paragraph{Preservation of natural Hamiltonians.}
Importantly, the isometric reductions above preserve the class of \emph{natural} frustration-free spin Hamiltonians: that is, the mapping $h_{a,u} \mapsto R_{u:uv}\herm h_{a,u} R_{u:uv}$ does not decrease the rank of the interaction on $\ens{a,u}$, and does not map the orthocomplement of the kernel to a space of the form $\ket{\phi} \otimes \mathbb C^2$ for any $\ket{\phi}$.

\paragraph{Reduction to the symmetric subspace.}
As a consequence of the previous feature, we may use the isometric reductions to map any ``natural'' frustration-free Hamiltonian to a ``complete homogeneous'' Hamiltonian $H_c$ on a system $V_c$, in which the non-zero terms $h_{a,b} = \ket{\beta_{a,b}}\bra{\beta_{a,b}}$ are supported on entangled states $\ket{\beta_{a,b}}$.
We show that the kernel of such a Hamiltonian has small dimension, and is spanned entirely by product states.
For an arbitrary spin $a \in V_c$, we may let $L_a = \idop$ and define a family of operators $L_v$ satisfying
\begin{align}
	\label{eqn:localMappingToAsymmetricState}
		\bra{\beta_{a,v}}
	\propto
		\bra{\Psi^-}_{a,v} \bigl( \idop_a \ox L_v \bigr) 
\end{align}	
for spins $v \in V_c$ and operators $\bra{\beta_{a,v}}$.
One then finds that $C = \bigotimes_{v \in V_c} L_v$ is a linear isomorphism (not necessarily an isometry) from the subspace $\Symm(\cH\sox{V_c})$ of symmetric states to the ground space of $H_c$.
(This isomorphism is also noted in Ref.~\cite[Section III A]{RandomQSat} 
for generic Hamiltonians with rank-1 interactions.)
As $\Symm(\cH\sox{V_c})$ is spanned by uniform superpositions $\ket{W_k}$ of standard basis states having Hamming weight $0 \leqslant k \leqslant n_c = |V_c|$, we have $\dim  \Symm(\cH\sox{n_c}) = n_c+1$.
This subspace may also be spanned by product state vectors $\ket{\alpha_0}\sox{n_c}\!,  \ldots,  \ket{\alpha_{n_c}}\sox{n_c}$ for any set of $n_c+1$ pairwise independent state vectors $\ket{\alpha_j} \in \cH$.
Thus, any complete homogeneous (natural) Hamiltonian $H_c$  has a ground space spanned by a family of vectors
\begin{align}
	\label{eqn:productsSpanKerHc}
		\ket{\Psi_j}
	=&
		\bigotimes_{v \in V_c} \bigl( L_v \ket{\alpha_j} \bigr)
	=
		C \ket{\alpha_j}^{\otimes n_c}
	,
\end{align}
for some $\ket{\alpha_0}, \ldots, \ket{\alpha_{n_c}} \in \cH$ as described.

Coupled with the tree-tensor structure of the isometric reductions, this characterization of the ground space of (natural) complete homogeneous spin Hamiltonians has the following consequences for (natural) frustration-free Hamiltonians:

\begin{obs}
	\label{obs:ffreeSample}
	For any frustration-free spin Hamiltonian $H_U$, any constant $k$, and for $k$-local operators $A$,  $\langle A \rangle$ can be efficiently computed with respect to ground states of $H$.
\end{obs}
Let $H_c$ be a complete homogeneous Hamiltonian obtained by isometric reduction of an unfrustrated Hamiltonian $H_U$, and let $n_c$ be the number of spins on which $H_c$ acts.
Consider a $k$-local operator $\tilde A$.
As $\ker(H_c)$ is spanned by some collection of product vectors $\ket{\Psi_0} = C \ket{\alpha_0}^{\otimes n_c}$, \ldots, $\ket{\Psi_{n_c}} = C \ket{\alpha_{n_c}}^{\otimes n_c}$ as in Eq.~\eqref{eqn:productsSpanKerHc}, we can 
efficiently compute the restriction of $\tilde A$ to $\ker(H_c)$ by evaluating the matrix
\begin{align}
		\label{eqn:skewRestriction}
		W(\tilde A)
	=&
		\sum_{j,k = 0}^{n_c} \ket{j}\bra{\Psi_j} \tilde A \ket{\Psi_k}\bra{k}
\end{align}
followed by a  suitable transformation.
Specifically, consider the operator $B = W(\idop)$; we have $B = U \Delta U^\dagger$ for some $U$ unitary and $\Delta$ positive and diagonal. We find
\begin{align}
			\Delta^{-1/2}  U^\dagger \sum_{j = 0}^{n_c}  \ket{j}\bra{\Psi_j}
		=
			\sum_{j = 0}^{n_c}  \ket{j}\bra{\Phi_j}	,	
\end{align}
for some orthonormal basis $\ket{\Phi_0}, \ldots, \ket{\Phi_{n_c}}$ of $\ker(H_c)$; thus, the restriction of 
$\tilde A$ to $\ker(H_c)$ 
with respect to the basis of states $\ket{\Phi_j}$ may be computed as
\begin{align}
		\bar A
	\;=&\;
		\Delta^{-1/2} U^\dagger W(\tilde A) U  \Delta^{-1/2}	 .
\end{align}
(For $\tilde A$ consisting of a single  $k$-spin term, the inner products of Eq.~\eqref{eqn:skewRestriction} consist of a product of constant-dimensional inner products; for $\tilde A$ a sum of multiple terms we extend linearly.)
Let $T: \ker(H_c) \to \ker(H_U)$ be the network of isometric reductions: then, by considering operators $\tilde A = T^\dagger A T$, we may compute the restriction $\bar A$ of such operators $A$ to the ground space of $H_U$.
We may then efficiently compute estimates using such (polynomial-size) matrices.

\begin{obs}
	The ground states of any frustration-free spin Hamiltonian $H_U$ on a lattice obey an entanglement area law.
\end{obs}

For any contiguous subsystem $A$ containing $a$ spins, we may reduce the Hamiltonian $H_U^{(A)}$ acting internally on $A$ --- by a tree-tensor isometry $T_A$ acting on $A$ alone, and inducing constraints as described by Eq.~\eqref{eqn:constraintInduction} --- to obtain a complete homogeneous Hamiltonian $H_c^{(A)}$, acting on at most $a$ spins.
The ground space of $H_c^{(A)}$ has dimension at most $a+1$; as 
the ground space of $H_U^{(A)}$ is an isometric image of that of $H_c^{(A)}$, the same is true of $H_U^{(A)}$.
As $H_U$ is frustration-free, any ground state of $H_U$ is also a ground state of $H_U^{(A)}$; it then follows that the Schmidt measure~\cite{SM} of $\ket{\Psi}$ with respect to the bipartition $V = A \union (V \setminus A)$ is at most $\log(a+1)$.
For a lattice with a well-defined dimension, we may obtain an area law for arbitrary subsystems of the spin lattice by summing this logarithmic bound over the number of distinct connected components.

\textit{``Almost'' frustration-free Hamiltonians.} We now leave the rigorous exact setting and turn to the 
physically interesting observation that the above ground state manifold can be used to grasp
approximately frustration-free models. Indeed, Observation~\ref{obs:ffreeSample} in particular suggests a 
variational approach to estimating ground energies for Hamiltonians $H$ for which
\begin{align}
		H	= H_U + \lambda H_F
\end{align}
for some $\lambda \ll 1$, where $H_U$ is frustration-free but the Hamiltonian $H$ itself is frustrated.
For such Hamiltonians $H$, if no phase transition is encountered,
the eigenvalues and eigenstates may differ little from those of the unfrustrated Hamiltonian $H_U$ (for 
related bounds on eigenvalues, see, e.g., Ref.\ \cite{BravyiTopo}). If the lowest $k$ 
eigenvectors (for some suitable $1 \le k \le n+1$) have a sufficiently 
high overlap with the lowest $k$ eigenvectors of $H_U$, we may approximately sample from the 
low-energy eigenvectors of $H$ by restricting to the kernel of $H_U$, using the efficient algorithm above.
In particular, as this procedure is variational, estimates obtained in this way for the ground state energy of $H$ are guaranteed to be upper bounds.

\begin{obs}
	\label{obs:almostffreeSample}
	The ground state manifolds of frustration-free Hamiltonians serve as an ansatz class for almost frustration-free models.
\end{obs}

We may also consider additional improvements to this estimation ansatz, in which less information about the ``frustrating'' component $\lambda H_F$ of the Hamiltonian is lost than in the tree-tensor renormalization procedure for the frustration-free model $H_U$.
To obtain better estimates --- and to extend these techniques to the case where $\lambda$ may be significantly large --- we may consider \textup{(a)}~\emph{partial} reductions by tree-tensor networks, having many free 
input spins; \textup{(b)}~isometries depending on the terms of the perturbed Hamiltonian $H$ 
(rather than those of $H_U$); and \textup{(c)}~using additional variational approaches.
We describe these ideas below.

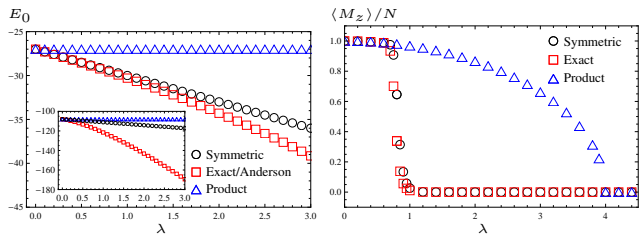
\begin{figure}[t]
\begin{center}
	\unitlength0.56mm
	\begin{picture}(70,50)
	\put(0,0){\includegraphics[width=4.1cm]{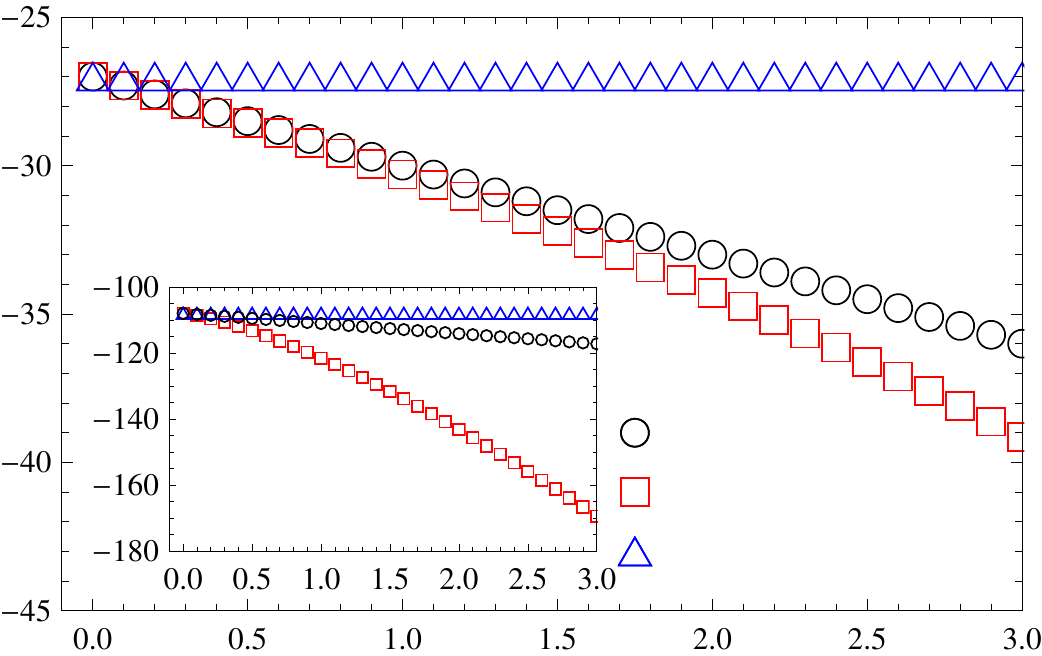} }
	\put(47,14.5){\tiny Symmetric}
	\put(47,10.5){\tiny Exact/Anderson}
	\put(47,6.5){\tiny Product}
	\put(35,-3){\tiny$\lambda$}
	\put(0,48){\tiny $E_0$}
	\end{picture}
	\hspace{2mm}
	\begin{picture}(70,50)
	\put(0,0){\includegraphics[width=4.1cm]{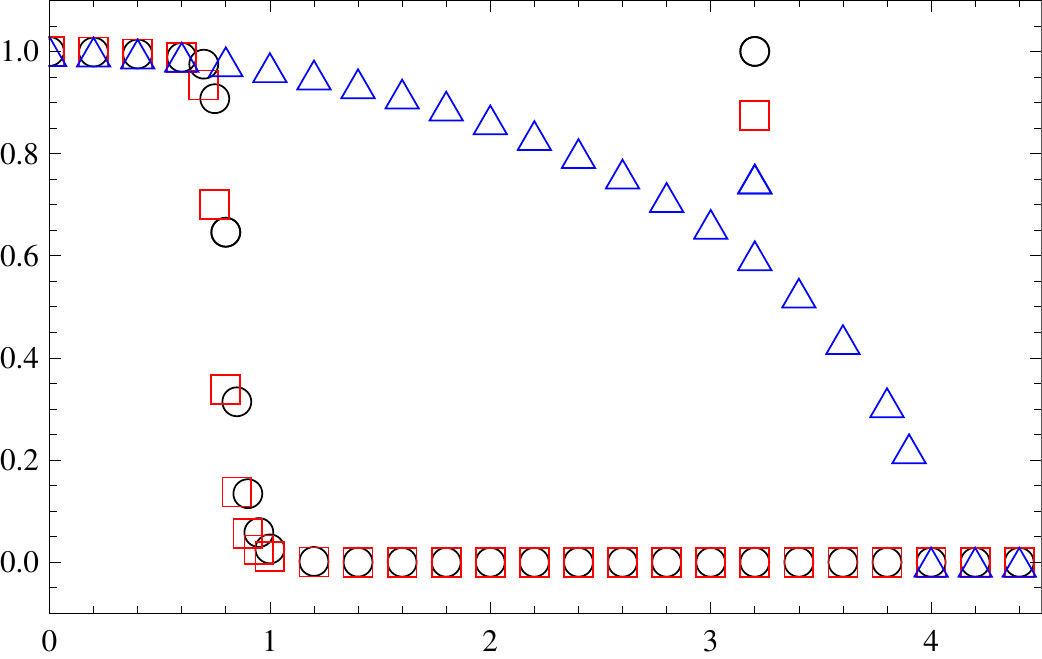}}
	\put(55.5,41.5){\tiny Symmetric}
	\put(55.5,37.2){\tiny Exact}
	\put(55.5,32.8){\tiny Product}
	\put(35,-3){\tiny $\lambda$}
	\put(0,48){\tiny ${\langle M_z\rangle}/N$}
	\end{picture}
  \end{center}

\caption{Left: Ground-state energy for XXZ-model on a trigonal lattice on a $3\times3$ torus, 
$h_{i,j}=-X_iX_j-Y_iY_j-(1-\lambda)Z_iZ_j$, by symmetric subspace estimate 
compared to product state ansatz and exact diagonalization. The inset shows the 
same model on a $6\times6$ torus where exact solution is not feasible and, therefore, is replaced by an Anderson lower bound.
Right:   Magnetization in z-direction for Ising model in a transverse field on a $4\times4$ torus,
$h_{i,j}=-Z_iZ_j$, $h_i=-\lambda X_i$, by 
symmetric subspace estimate compared to Gutzwiller  
mean-field approximation (product state ansatz) 
and exact diagonalization.}\label{Figure}
\end{figure}

In the case that the ground state manifold of $H_U$ is contained in a subspace $\mathcal K$ which is spanned by product states and has ``small'' dimension (i.e., polynomial rather than exponential in the system size, as in 
the case of the symmetric subspace for ferromagnetic Ising or XXX models), we may indeed forgo isometric reductions entirely, and estimate the ground energy of $H$ by considering the restriction of $H$ to $\mathcal K$ using the techniques described for Observation~\ref{obs:ffreeSample}.
In contrast to a full tree-tensor contraction, this has the advantage of yielding exact results for the frustration-free case $\lambda = 0$ while retaining more information about the ``frustrating'' component $H_F$.
The resulting estimate for the ground state energy will be a linear function of $\lambda$, whose value and first derivative with respect to $\lambda$ agree with that of the exact ground energy at $\lambda = 0$.
For ``small'' $\lambda$ and modest system sizes, this may yield a good estimate of 
the ground state energy of $H$: see Fig.~\ref{Figure} \cite{Note2}.

To obtain estimates which account for spatially decaying correlations, we may perform a \emph{partial} tree-tensor reduction with a small number of contraction layers, and sample with respect to a subspace $\mathcal K$ as described above.
In each layer, we may fix a collection of (non-intersecting) adjacent site pairs $\{a,b\}$ to contract, and for each such pair $\{a,b\}$ apply some isometric contraction as described in Eq.~\eqref{eqn:isometricContraction}.
However, rather than apply the reductions which 
would be suggested by the frustration-free Hamiltonian $H_U$, we may use isometries
\begin{align}
		Q_{a:ab} = \ket{\psi^{0}_{a,b}}\bra{0} + \ket{\psi^1_{a,b}}\bra{1} ,
\end{align}
where $\ket{\psi^0_{a,b}}, \ket{\psi^1_{a,b}}$ are the lowest energy eigenvectors 
of $\eta_{a,b}$, where $\eta_{a,b}$ is the partial trace of $H$ with respect to all sites other than $a$ and $b$.
Given a tensor network $T_1$ consisting of a product of such two-site operators $Q_{a:ab}$, we may then consider the spin model given by $H' = T_1^\dagger H T_1$, and estimate the ground energy of $H'$ with respect to a low-dimension subspace or another isometric contraction; this yields an upper bound on the ground energy of the original Hamiltonian $H$.

As expected, the symmetric subspace estimate works best for almost-unfrustrated systems.
The region where the results are accurate is larger in small and medium size systems, where the improvement compared to mean-field is most significant: see Fig.~\ref{Figure}. For the XXZ-model, applying isometric reductions on disjoint pairs of nearest neighbors improves the approximation dramatically (versus symmetric states) for ``strongly'' frustrated interactions, as 
they allow for the selection of the most important two-site subspaces \cite{Note0}.

The approximation technique presented here can also be combined with any variational method.
One simple idea is to apply single-site unitaries to the Hamiltonian before performing the restriction to the symmetric subspace.
One may then vary the (different) unitaries applied on the sites to minimize the energy, 
in order to find a better approximation to the ground state. This can lead to a notable improvement, 
especially when antiferromagnetic effects are non-negligible. By construction, this approach performs at least as well as a product state ansatz to which it reduces for $\lambda\to\infty$ \cite{Note}. 

\textit{Summary.} In this work, we have introduced a class of spin models that can be {\it completely solved}: The entire 
ground state manifold can be explicitly given and parametrized by an entire symmetric subspace under a tensor network,
so a ``tensor network with an input''. 
This class of models is hence expected to constitute a rich playground of exploring ideas on quantum lattice models,
complementing work that exemplifies how computationally {\it difficult} it can be to even approximate ground state
energies. In a sense, the considered models can 
also be viewed as the {\it parent Hamiltonian} of the network, in a converse approach taken for tree tensor networks in
Ref.\ \cite{Giovannetti}. We always find area laws for the entanglement entropy, 
hence establishing a new large class of models beyond free systems
for which an area law can be proven to hold. It is the hope that this work stimulates further research on 
models for which tensor networks not only arise as computational  but as essentially analytical tools.

{\it Acknowledgements}
We acknowledge discussions with D.\ Gross, S. Michalakis, and T.\ Barthel and support 
by the EU (QESSENCE, MINOS, COMPAS) and the EURYI scheme. 



\begin{thebibliography}{99}

\bibitem{Area1}
	J.\ Eisert, M.\ Cramer, and M.~B.\ Plenio, Rev.\ Mod.\ Phys. {\bf 82}, 277 (2010).
			
\bibitem{Area2}		
	M.\ Srednicki, 
	Phys.\ Rev.\ Lett.\ {\bf 71}, 666 (1993).
	
\bibitem{Area3}		
	K.~M.~R.\ Audenaert, J.\ Eisert, M.~B.\ Plenio, and R.~F.\ Werner,
	Phys.\
	Rev.\ A {\bf 66}, 042327 (2002);		
	G.\ Vidal, J.~I.\ Latorre, E.\ Rico, and A.\ Kitaev, 
	Phys.\ Rev.\
	Lett.\ {\bf 90}, 227902 (2003);
	B.-Q.\ Jin and V.\ Korepin,  J.\ Stat.\ Phys.\ {\bf 116}, 79 (2004);
	P.\ Calabrese and J.\ Cardy, 
	J.\ Stat.\ Mech.\ P06002 (2004);
	M.~B.\ Plenio, J.\ Eisert, J.\ Dreissig, and M.\ Cramer,	
	Phys.\ Rev.\ Lett.\ {\bf 94}, 060503 (2005);
	M.~B.\ Hastings, JSTAT P08024 (2007).

\bibitem{Scholl}
	F.\ Verstraete, J.\ I.\ Cirac, and V.\ Murg,
	Adv.\ Phys.\ {\bf 57}, 143 (2008);
	U.\ Schollwoeck,
	Rev.\ Mod.\ Phys.\ {\bf 77}, 259 (2005).
	
\bibitem{Power}		
	D.\ Aharonov, D.\ Gottesman, S.\ Irani, and J.\ Kempe,
	Comm.\ Math.\ Phys.\ {\bf 287}, 41 (2009).

\bibitem{Bravyi06}
	S.~Bravyi, quant-ph/0602108.

\bibitem{Farhi}
	R.\ Movassagh, E.\ Farhi, J.\ Goldstone, D.\ Nagaj, T.\ J.\ Osborne, and P.\ W.\ Shor,
	arXiv:1001.1006.

\bibitem{SM}
	J.\ Eisert and H.\ J.\ Briegel, Phys.\ Rev.\ A {\bf 64},  022306 (2001).

\bibitem{Note0}	
	We similarly accumulate any single-spin contributions in $H'_U$ which arise from single-spin contributions in $H_U$.
	
\bibitem{Note}
	While we effectively ignore single-spin terms in our analysis, 
	such terms do impose constraints on the ground state manifold, 
	and thus on whether the Hamiltonian is frustration free.
	However, these may be accounted for with little difficulty.

\bibitem{MERA}
	G.\ Vidal, Phys.\ Rev.\ Lett.\ {\bf 99}, 220405 (2007).

\bibitem{RandomQSat}
	C.~R.\ Laumann, A.~M.\ L{\"a}uchli, R.\ Moessner, A.\ Scardicchio, and S.~L.\ Sondhi,
	arXiv:0910.2058.

\bibitem{BravyiTopo}
	S.\ Bravyi, M.\ B.\  Hastings, and S.\ Michalakis,
	arXiv:1001.0344.

\bibitem{Note2}	
	However, for systems whose ground states have very localized correlations, 
	estimating with respect to a subspace spanned by a small number of product 
	states may scale poorly for large systems.
	For instance, when the variational set $\mathcal K$ is the symmetric subspace 
	$\Symm(\mathbb C^N)$, the very symmetry of states in $\mathcal K$ 
	entails that correlations do not decay spatially but rather are constant.

\bibitem{Note0}	
	The approach taken here is also expected to work for slightly
	frustrated Hamiltonians reminiscent of {\it Shastry-Sutherland-type models} \cite{SS81} on
	cubic lattices, where an additional bond along the main diagonal renders the
	model frustrated.

\bibitem{SS81}	
	B.\ S.\ Shastry, B.\ Sutherland, Physica B \textbf{108}, 1308 (1981).

\bibitem{Note}				
	One could also perform more sophisticated quantum 
	circuits, such as constant-depth quantum circuits or partial 
	MERA reductions: there will then exist a trade-off between accuracy of the simulation and the efficiency of the procedure.

\bibitem{Giovannetti}
	P.\ Silvi, V.\ Giovannetti, S.\ Montangero, M.\ Rizzi, J.\ I.\ Cirac, and R.\ Fazio,
	arXiv:0912.0466.
	
\end{thebibliography}
\end{document}